\documentclass[
    10pt,
    journal,
    compsoc,
    showpacs,
    reprint,
    twocolumn,
    superscriptaddress,
    floatfix,
    amsmath
]{IEEEtran}

\ifCLASSOPTIONcompsoc
  \usepackage[nocompress]{cite}
\else
  \usepackage{cite}
\fi

\usepackage[font=small,labelfont=bf]{caption}
\usepackage{braket}

\usepackage{graphics}
\usepackage{multirow}
\usepackage{dcolumn}
\newcolumntype{.}{D{.}{.}{-1}}

\usepackage{epsfig}
\usepackage{graphicx}
\usepackage{epstopdf}
\usepackage{hyperref}
\usepackage{verbatim}
\usepackage[english]{babel}

\usepackage{color}
\usepackage{listings}
\definecolor{codegreen}{rgb}{0,0.6,0}
\definecolor{codegray}{rgb}{0.5,0.5,0.5}
\definecolor{codepurple}{rgb}{0.58,0,0.82}
\definecolor{backcolour}{rgb}{0.98,0.98,0.98}
 
\lstdefinestyle{mystyle}{
    backgroundcolor=\color{backcolour},   
    commentstyle=\color{codegreen},
    keywordstyle=\color{magenta},
    numberstyle=\tiny\color{codegray},
    stringstyle=\color{codepurple},
    basicstyle=\ttfamily\scriptsize,
    breakatwhitespace=false,         
    breaklines=true,                 
    captionpos=b,                    
    keepspaces=true,                 
    numbersep=5pt,                  
    showspaces=false,                
    showstringspaces=false,
    showtabs=false,                  
    tabsize=2
}
 
\makeatletter
\long\def\@IEEEtitleabstractindextextbox#1{\parbox{0.922\textwidth}{#1}}
\makeatother

\begin{document}

\title{Continuous evaluation of the performance of cloud infrastructure for scientific applications}

\author{
    \IEEEauthorblockN{
        Mohammad Mohammadi, Timur Bazhirov}\\
        \IEEEauthorblockA{
            Exabyte Inc., San Francisco, California 94103, USA\\
        }
}

\IEEEtitleabstractindextext{%
    
    \begin{abstract}
    
        Cloud computing recently developed into a viable alternative to on-premises systems for executing high-performance computing (HPC) applications. With the emergence of new vendors and hardware options, there is now a growing need to continuously evaluate the performance of the infrastructure with respect to the most commonly-used simulation workflows. We present an online ecosystem and the corresponding tools aimed at providing a collaborative and repeatable way to assess the performance of the underlying hardware for multiple real-world application-specific benchmark cases. The ecosystem allows for the benchmark results to be stored and shared online in a centrally accessible database in order to facilitate their comparison, traceability and curation. We include the current up-to-date example results for multiple cloud vendors and explain how to contribute new results and benchmark cases.
    
    \end{abstract}
    
}

\maketitle


\section{Introduction}
\label{sec:introduction}

    Cloud computing now represents a viable alternative to on-premises hardware for high-performance scientific applications. Although the feasibility and advantages of cloud computing for high-performance computing (HPC) workloads were debated for over a decade \cite{2009-napper, 2011-iosup, 2010-jackson-cholia-lbl-cloud-con, 2011-cloud-computing-magellan-report}, the recent advancements in the field make it into a competitive and cost-effective solution to run compute-intensive parallel workloads for a large variety of models and application areas \cite{2008-deelman, 2008-evangelinos, 2007-keahey, 2008-keahey, 2009-keahey, 2010-li, 2010-ramakrishnan, 2010-rehr, 2012-sanjay, Sadooghi, Connor-2018, accelerating-discovery-2012, cloud-simulations-on-google}. In our previous studies \cite{2016-exabyte-aps-abstract, 2018-exabyte-accessible-CMD, 2018-exabyte-binary-compounds, exabyte2018hp3c, 2018-exabyte-phonon} we demonstrated that HPC in the cloud is ready for a widespread adoption and can provide a viable and cost-efficient alternative to capital-intensive on-premises hardware deployments for large-scale high-throughput and distributed memory calculations. 
    
    
    With the ever-increasing adoption of cloud computing, the emergence of new vendors and computing hardware, there is a growing need for continuous evaluation of the infrastructure performance for the real-world workloads. However, the lack of intuitive and collaborative tools have made such an assessment challenging. There exist multiple previous works in the field either, however they usually focus on generic computing and leave out real-world application use cases and lack a collaborative and systematic ecosystem to share and manage the results \cite{Tailbench-2016, performance-modeling-2016, micro-benchmarks-2018, iaas-benchmark-2015, cloudsuite, spec}.
    
    We hereby present the concept and the associated software tools for an online benchmarking ecosystem able to continuously and collaboratively assess the performance of computing hardware for real-world scientific applications. The aim of the ecosystem is three-fold: to assist community in choosing the best hardware for HPC applications, to allow cloud vendors to identify the bottlenecks and improve their services, and help HPC application developers to identify and address implementation-related challenges accordingly.
    
    The ecosystem includes ExaBench suite \cite{exabyte-benchmarks-repo}, an open-source modular and extensible software tool able to facilitate the performance assessment of computing systems and a centrally accessible collaborative online repository to store and manage the results. In the following, we outline the ecosystem and its operations, provide example results, and explain how to contribute to its further development. 

    \begin{figure}[ht!]
    \includegraphics[width = 0.48\textwidth]{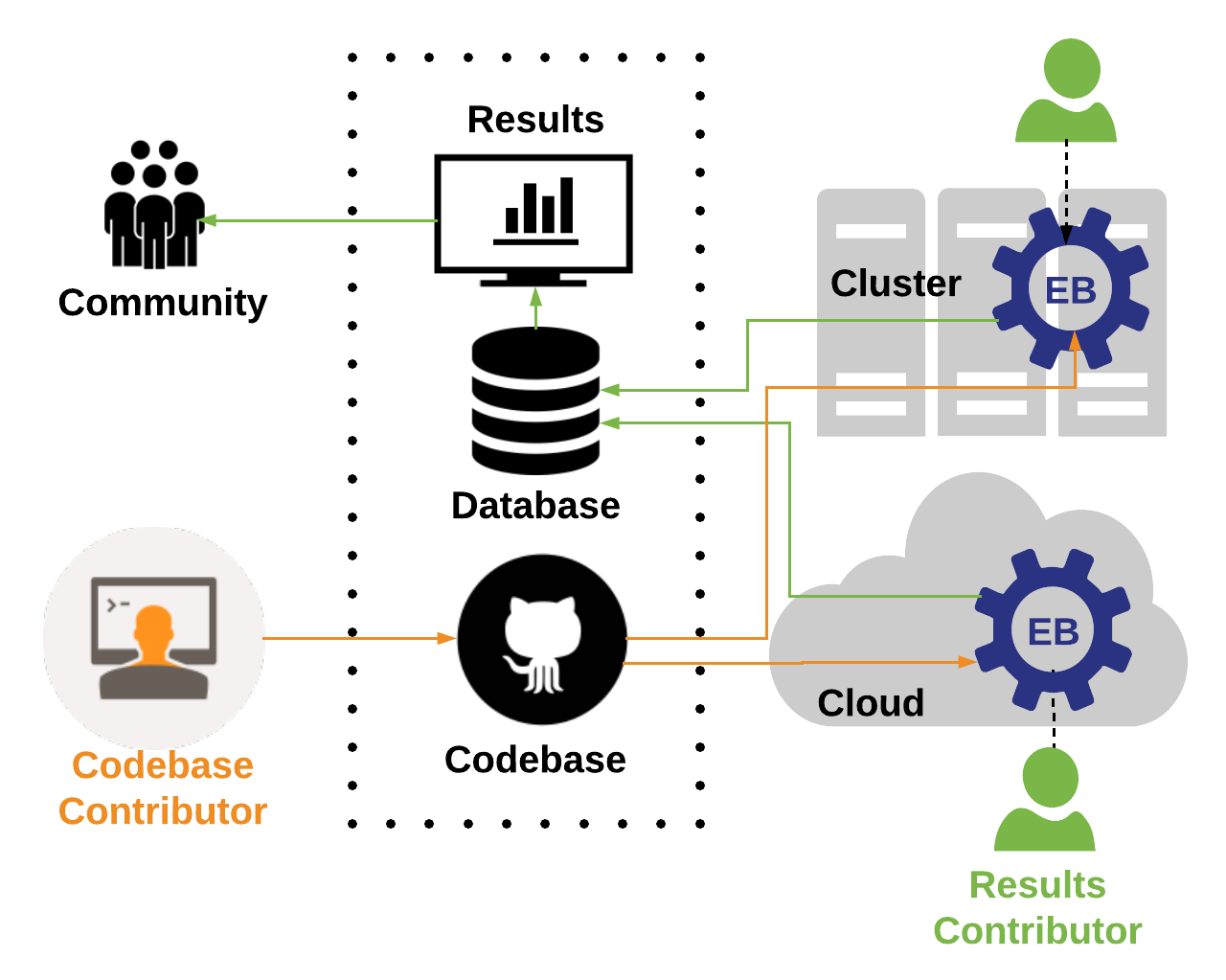}
    \caption{ 
        \label{exabench-figure}
        Schematic representation of the online ecosystem presented in this manuscript. The three main components - Open-source codebase, Database of results and their online visual representation, are outlined in the middle. "EB" denotes the ExaBench tool. "Cluster" refers to on-premises computing clusters. "Cloud" denotes the public/private cloud systems. Two types of contributors - to codebase (in orange) and to results (green) are shown. Codebase contributors help extend the test cases. Results contributors run ExaBench tool and publish the results to the centrally accessible database. The results are available to the wider community.
        }
    \end{figure}


\section{Ecosystem}
\label{sec:ecosystem}
    
    We view the ecosystem as an online platform allowing multiple people to collaboratively evaluate the performance of computing hardware for compute-intensive applications. 
    
    \subsection{Components}
    \label{sec:ecosystem-components}
    
        We identify the following components:
    
        \begin{itemize}
        
            \item{\textbf{ExaBench}}, an open-source modular software tool to facilitate the performance assessment of computing systems. The tool supports multiple benchmark cases to evaluate the performance of scientific applications.
            
            \item{\textbf{Results database}}, a centrally accessible repository to store the results in order to facilitate their comparison, traceability and curation.
            
            \item{\textbf{Results page}}, an online resource presenting the results in a visual manner.
            
            \item{\textbf{Sites}}, or physical location with unique identifiers where the benchmarks are executed.
            
            \item{\textbf{Contributors}}, who use the ExaBench tool to submit the benchmark results to the central database and/or contribute to the ExaBench source code by adding support for new benchmark cases and metrics,.
        
            \item{\textbf{Community}}, the broader set of users and interested parties.
        
        \end{itemize}
    
        Schematic representation of is available in Fig.~\ref{exabench-figure}. 
        
    \subsection{Operations}
    \label{sec:ecosystem-components}
    
        When considering the functions of the ecosystem, we envision the following process. Sites' administrators install ExaBench tool from the source code available online and developed and maintained continuously by the codebase contributors (including the authors of this manuscript, originally). Benchmarks are executed on the underlying hardware and their results are stored automatically in the database in a certain format. In order to evaluate the performance the benchmark cases are executed with entirely equivalent setups for the number of nodes and processors per node configurable in the tool. The execution time is used to evaluate the performance \cite{computer-architecture}. The cases are designed to be compact so that they can be executed within a reasonable timeframe on sites with different hardware configurations. The results are stored in an centrally accessible database for the community to analyze the efficiency of the computing systems.
    

\section{ExaBench}
\label{sec:exabench}

    Exabyte benchmarks suite (ExaBench) is an open-source modular and extensible software tool written in Python aimed to help the scientific and engineering communities to asses the performance of cloud- and on-premises systems. The suite consists of three main components, benchmarks, metrics, and results, outlined in the following sections.

    \subsection{Benchmarks}
    
        "Benchmarks" package implements benchmark cases for real-world scientific applications. Each application is introduced as a sub-package containing multiple benchmark cases, templates, and input files to cover different application execution patterns. The cases are implemented in an object-oriented modular way representing a class with methods to prepare and execute the simulation(s) and extract its results. The following summarizes the currently supported benchmark types:
        
        \begin{itemize}
            \item High-Performance Linpack (HPL)
            \item Vienna Ab-initio Simulation Package (VASP \cite{VASP})
            \item GRoningen MAchine for Chemical Simulations (GROMACS \cite{GROMACS})
        \end{itemize}
    
    \subsection{Metrics}
    
        "Metrics" package implements the metrics used to analyze the results. The list of currently supported metrics is given below. The term "performance gain" is referred to the ratio of performance in giga-FLOPS (GFLOPS) for a given number of nodes to the performance for a single node.
    
        \begin{itemize}
        \item{\textbf{Speedup ratio}}: The ratio of the Performance Gain defined above for a given number of nodes to the ideal speedup. This metric is used for HPL benchmark cases to estimate the extent to which HPL calculations can be efficiently scaled.
                
        \item{\textbf{Speedup}}: The inverse of total runtime in seconds for a benchmark case. The metric is used to understand how quickly application-specific benchmark cases (e.g. VASP) can be executed.
    
        \item{\textbf{Performance per core}}: The ratio of performance in giga-FLOPS (GFLOPS) to the total number of cores used by the benchmark case.
        \end{itemize}
    
    \subsection{Results} 
        
        "Results" package implements the necessary handlers to store the results. When the benchmarks are executed, their results are stored and shared online in a centrally accessible and collaborative repository \cite{google-spreadSheet}. Readers should note that the data stored there is preliminary/raw and so might not be accurate. Nevertheless, it allows to automate the process and minimizes human error. We automatically generate the charts for the metric explained above to compare the sites. Each point in the graphs is the average of existing results for the specific site and configuration as the benchmarks may be executed multiple times on a site.


\section{Example Results}
\label{sec:example-results}

    Below we present some example results of benchmarks performed using "ExaBench" tool and available online as part of the ecosystem. For more details the readers are referred to the full explanation available in \cite{exabyte-documentation}. 
    
    \subsection{High-performance Linpack}

\begin{figure}[ht!]
      \includegraphics[width = 0.48\textwidth]{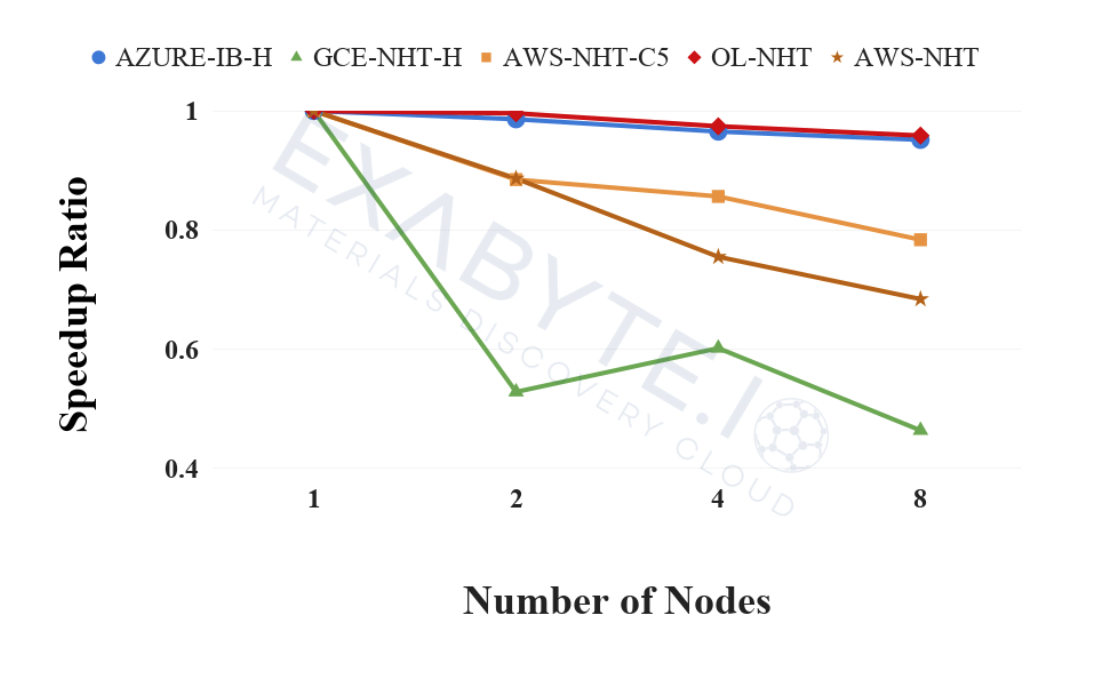}
      \caption{
        \label{speedup-ratio-figure}
        Speedup Ratio vs Number of Nodes. Speedup ratio for 1,2,4 and 8 nodes are investigated and given by the points. Lines are drawn to guide the eye. The legend is as follows: AWS-NHT - Amazon Web Services with hyper-threading disabled; AWS-NHT-C5 - same as AWS-NHT with C5 instances; AZ-IB-H - Microsoft Azure Infiniband-interconnected H16r VMs; OL-NHT - Oracle Cloud BM.HPC2.36 instances with hyper-threading disabled; GCE-NHT-H - Google Compute Engine n1-highcpu-64 machines with hyper-threading disabled, Haswell platform.
        }
    \end{figure}

    We present a comparison of the speedup ratios for High-performance Linpack benchmark In Fig.~\ref{speedup-ratio-figure}. We consider 4 cloud computing vendors: Amazon Web Services (AWS) with c4 (default and c5 instance types, Microsoft Azure (AZ), Oracle Cloud (OL), and Google Compute Engine (GCE). The speedup ratio metric is obtained from the results of HPL benchmark running on 1,2,4, and 8 nodes with hyper-threading disabled. As it can be seen, Oracle and Microsoft Azure exhibit better scaling because of the low-latency interconnect network.
    
    \subsection{Vienna ab-initio simulation package}
    
\begin{figure}[ht!]
      \includegraphics[width = 0.48\textwidth]{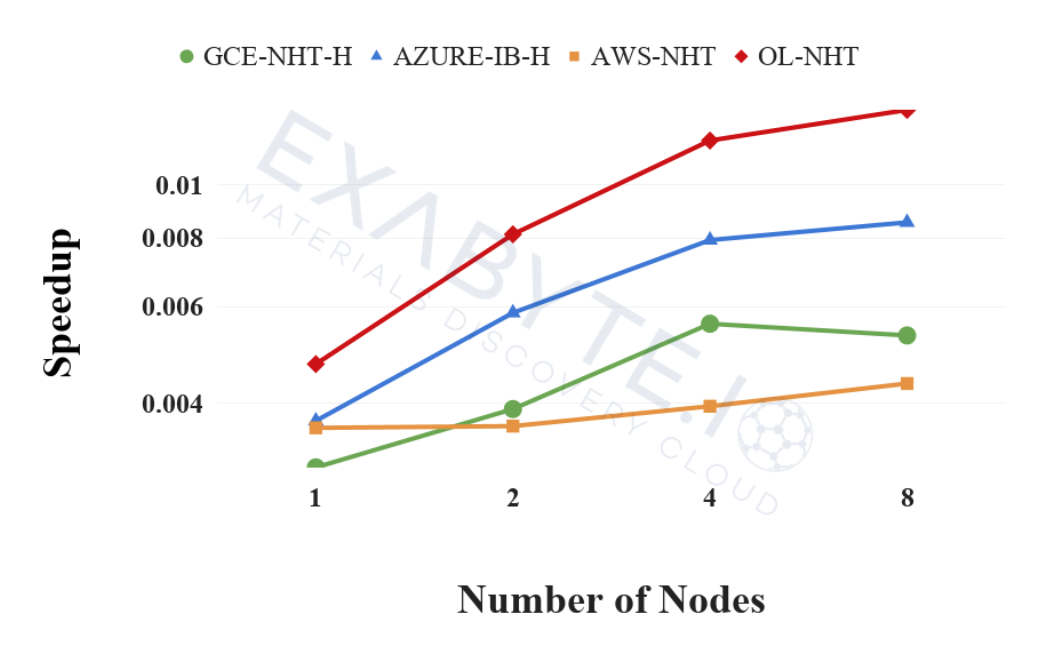}
      \caption{
        \label{vasp-elb-figure}
        Speedup vs Number of Nodes for Vienna ab-initio simulations package, parallelization over electronic bands. Speedup for 1,2,4 and 8 nodes are investigated and given by the points. Lines are drawn to guide the eye. The legend is as follows: AWS-NHT - Amazon Web Services with hyper-threading disabled; AZ-IB-H - Microsoft Azure Infiniband-interconnected H16r VMs; OL-NHT - Oracle Cloud BM.HPC2.36 instances with hyper-threading disabled; GCE-NHT-H - Google Compute Engine n1-highcpu-64 machines with hyper-threading disabled, Haswell platform.
        }
    \end{figure}

    A comparison of the speedups for Vienna ab-initio simulation package (VASP) is given in Fig.~\ref{vasp-elb-figure}. We show results for the test case involving the parallelization over the electronic bands for a large-unit-cell material (refered to as "VASP-ELB"). We use version 5.3.5 with the corresponding set of atomic pseudopotentials. The goal of this benchmark is to estimate the extend to which a VASP calculation can be efficiently scaled in a distributed memory execution scenario. As it can be seen, the vendors with higher CPU clockspeed and low-latency interconnect network perform better.
    
\section{Contribution}
\label{sec:Contribution}

    We embrace the open-source online character of the ecosystem presented here and encourage collaborative contributions. The ecosystem can be further extended in two ways, by contributing to the results or extending the codebase. 
    
    \subsection{Contributing to the results}
        
        In order to contribute to the results one should configure, and execute the benchmarks and send the results to the central database, all with the help of ExaBench tool. Readers are referred to the comprehensive explanation about the installation, configuration, and operation of the tool available online inside the corresponding GitHub repository \cite{exabyte-benchmarks-repo}.
    
    \subsection{Extending ExaBench}
        
        In order to extend the source code with new cases and metrics, it is recommended to "fork" the repository and introduce the adjustments there. The changes in the fork can further be considered for merging into the repository as it is commonly used on GitHub. This process is explained in more details elsewhere online \cite{github-fork} and inside the ExaBench repository itself \cite{exabyte-benchmarks-repo}.


\section{Perspectives and Outlook}
\label{sec:perspectives-and-outlook}

    High-performance and parallel computing today is more important than ever due to the end of Moore's law in conventional semiconductor technology scaling. HPC is no longer a domain of highly specialized applications only. The latter still exist and are needed, but gradually become a minority. For that reason and in order to facilitate the timely and objective insights, the importance of a continuous collaborative performance assessment is strong today and will grow further in the future. Following the limited set of applications we incorporated into the ecosystem today, many more use cases in computational fluid dynamics, electronic design automation, drug discovery, computational chemistry, etc. can be introduced by extending the source code and contributing the results.

    We envision that the ecosystem will help the community to choose the optimal setup for running resource-intensive workloads, and let cloud vendors to improve their services in a competitive and transparent environment. We see how such an environment can lead to further democratization of HPC and its proliferation in the industrial research and development, which in turn will accelerate progress in the corresponding industries.


\section{Conclusion}
\label{sec:conclusion}

    In this manuscript we present an online ecosystem to study the efficiency and suitability of computing hardware for a variety of real-world high-performance computing applications. The ecosystem provides a collaborative, continuous and transparent evaluation of the cloud as well as on-premises infrastructures by automating the benchmarking process and storing the results in a centrally accessible repository available to the community. We showcase example results, and explain how to contribute the assessment of performance metrics for new sites, and how to extend the package further by adding new benchmark cases and metrics.

\bibliographystyle{IEEEtran}
\bibliography{ref}

\end{document}